\documentstyle[12pt,openbib]{article}
\begin{document}

\title{Einstein-Bianchi Hyperbolic System\\ for General Relativity\cite{ded}}

\author{Arlen Anderson, Yvonne Choquet-Bruhat\cite{CBadd},\\ and
James W. York Jr.\\
	{\it Department of Physics and Astronomy}\\
        {\it University of North Carolina}\\
	{\it Chapel Hill, NC 27599-3255}
          }

\date{Sep. 8, 1997}

\def\al{{\alpha}}
\def\be{{\beta}}
\def\ga{{\gamma}}
\def\gam{{\gamma}}
\def\de{{\delta}}
\def\la{{\lambda}}
\def\eps{{\epsilon}}
\def\th{{\theta}}
\def\sig{{\sigma}}
\def\om{{\omega}}

\def\Gam{{\Gamma}}
\def\bGam{{\bar\Gamma}}
\def\bbGam{\mbox{\boldmath $\bGam$}}

\def\barbg{{\bar{\bf g}}}
\def\bg{{\bf g}}

\def\bfe{\mbox{\boldmath $\bar e$}}

\def\bA{{\bf A}}
\def\bB{{\bf B}}
\def\bC{{\bf C}}
\def\bD{{\bf D}}
\def\bE{{\bf E}}
\def\bF{{\bf F}}
\def\bG{{\bf G}}
\def\bH{{\bf H}}
\def\bJ{{\bf J}}
\def\bK{{\bf K}}
\def\bS{{\bf S}}

\def\ba{{\bf a}}
\def\bb{{\bf b}}
\def\bd{{\bf d}}
\def\bbe{{\bf e}}
\def\bh{{\bf h}}
\def\bk{{\bf k}}
\def\bell{{\mathbf{\ell}}}
\def\bu{{\bf u}}

\def\bbeta{\mbox{\boldmath $\beta$}}
\def\bgam{\mbox{\boldmath $\gamma$}}
\def\brho{\mbox{\boldmath $\rho$}}
\def\bphi{\mbox{\boldmath $\phi$}}
\def\bpsi{\mbox{\boldmath $\psi$}}
\def\bxi{\mbox{\boldmath $\xi$}}
\def\bsig{\mbox{\boldmath $\sigma$}}

\def\cC{{\mathcal{C}}}
\def\cE{{\mathcal{E}}}
\def\cP{{\mathcal{P}}}

\def\d{{\partial}}
\def\dzeroh{{\hat\partial_0}}

\def\tr{{\mathrm{tr}}}
\def\trK{{\tr\bK}}

\def\Sup{{\mathrm{Sup}}}

\def\bgrad{{\bar\nabla}}
\def\bbgrad{{\mathbf{\bar\nabla}}}

\newcommand\beq{\begin{equation}}
\newcommand\eeq{\end{equation}}
\newcommand\beqa{\begin{eqnarray}}
\newcommand\eeqa{\end{eqnarray}}

\maketitle
\vspace{-11cm}
\hfill IFP-UNC-521

\hfill TAR-UNC-059

\hfill gr-qc/9710041
\vspace{10cm}
\begin{abstract}
By employing the Bianchi identities for the Riemann tensor in conjunction
with the Einstein equations, we construct a first order symmetric
hyperbolic system for the evolution part of the Cauchy problem of
general relativity.  In this system, the metric evolves at zero speed
with respect to observers at rest in a foliation of spacetime by
spacelike hypersurfaces while the curvature and connection propagate
at the speed of light.  The system has no unphysical characteristics,
and matter sources can be included.
\end{abstract}
\newpage

\section{Introduction}

\indent We consider the evolution part of the Cauchy problem in General 
Relativity\cite{cite1} as the time history of the two fundamental
forms of a spacelike hypersurface: its metric $\barbg$ and its
extrinsic curvature $\bK$.  On such a hypersurface, for example
an ``initial'' one, these two quadratic forms must satisfy four
initial value or constraint equations.  These constraints can be
posed and solved as an elliptic system by known methods that
will not be discussed here.  (See, for example, \cite{cite1}.)

The Ricci tensor of the spacetime metric can be displayed in a 
straightforward 3+1 decomposition giving the time derivatives of
$\barbg$ and $\bK$ in terms of the space derivatives of these 
quantities.  These expressions contain also the lapse and shift
functions characterizing the threading of the spacelike hypersurfaces
by time lines.  However, proof of the existence of a causal evolution
in local Sobolev spaces into an Einsteinian spacetime does not
result directly from these equations, which do not form a hyperbolic
system for arbitrary lapse and shift, despite the fact that their
characteristics are only the light cone and the normal to the time
slices\cite{cite2}.

In this paper we present in detail, using ``algebraic gauge'' or
a generalized harmonic time slicing condition
on the lapse function, a hyperbolic system for these geometric unknowns
that is based directly on the Riemann tensor, using the Bel\cite{cite7} 
electric and magnetic fields and inductions, and the full Bianchi
identities.  Its characteristics are the physical light cone and
the time direction orthogonal to the spacelike submanifolds
$M_t$ of the chosen slicing.  The spacetime curvature tensor is
among the quantities that are evolved explicitly.  Its propagation
is coupled to that of the connection, extrinsic curvature, and metric
of $M_t$ through relations introduced by H. Friedrich\cite{cite2},
who obtained an analogous vacuum system based on the Weyl tensor,
causal but with additional characteristics.
Our system constitutes a complete first order symmetric
hyperbolic system equivalent to the usual Einstein equations
including matter.
This new system, which was sketched briefly in \cite{cite3},
is completely equivalent to our previous one \cite{cite4,cite5,cite6}.
The space coordinates and the shift three-vector are arbitrary;
and, in this sense, the system is independent of a gauge choice.
In fact, we can also make the lapse function take arbitrary
values by introducing a given arbitrary function into its definition.
Indeed, our procedure for the lapse function is equivalent to 
specifying $\bar g^{-1/2} N$ as an arbitrary positive function of
the spacetime coordinates.

\section{Metric, Connection, and Curvature}

\indent It is convenient to write the metric on the hyperbolic spacetime
manifold $M_t\times R$ as
\beq
ds^2 = - N^2 (\th^0)^2 + g_{ij} \th^i \th^j,
\eeq
with $\th^0=dt$ and $\th^i= dx^i + \beta^i dt$, where $\beta^i$ is
the shift vector.  The cobasis $\th^\al$ satisfies
\beq
d\th^\al = - {1\over 2} C^\al\mathstrut_{\be\ga} \th^\be \wedge \th^\ga,
\eeq
with $C^i\mathstrut_{0j}= -C^i\mathstrut_{j0} = \d_j \be^i$ and all
other structure coefficients zero.  The corresponding vector basis
is given by $e_0 = \d_t - \be^j \d_j$, where $\d_t = \d/\d t$, 
$\d_i = \d/\d x^i$, and the action of $e_0$ on space scalars is the
Pfaffian or convective derivative
$$ e_0[f] = \d_0 f = \d_t f - \be^j \d_j f.$$

We shall assume throughout that the lapse function $N>0$ and the
space metric $\barbg$ on $M_t$ is properly Riemannian.  Note that
$\bar g_{ij} = g_{ij}$ and $\bar g^{ij} = g^{ij}$.  (An overbar
denotes a spatial tensor or operator.)

The connection one-forms are defined by
\beq
\om^\al\mathstrut_{\be\ga} = \Gam^{\al}\mathstrut_{\be\ga} + 
g^{\al\de} C^\eps\mathstrut_{\de(\be} g_{\ga)\eps} - 
{1\over 2} C^{\al}\mathstrut_{\be\ga},
\eeq
where $(\be\ga)= {1\over 2}(\be \ga + \ga \be)$, 
$[\be \ga] = {1\over 2}(\be \ga - \ga \be)$, and $\Gam$ denotes a
Christoffel symbol.  We adhere to the convention
\beq
\nabla_\al v^\be = \d_\al v^\be + v^\sig \om^\be\mathstrut_{\sig\al}
\eeq
for the spacetime covariant derivative $\nabla_\al$ and likewise for
the spatial covariant derivative $\bgrad_i$.  The Riemann 
curvature tensor is given by
\beq
\label{Riemann}
R^\al\mathstrut_{\be,\rho\sig} = \d_\rho \om^\al\mathstrut_{\be\sig}
-\d_\sig \om^{\al}\mathstrut_{\be\rho} +
\om^\al\mathstrut_{\la\rho} \om^\la\mathstrut_{\be\sig}
-\om^\al\mathstrut_{\la\sig} \om^\la\mathstrut_{\be\rho}
-\om^\al\mathstrut_{\be\la} C^{\la}\mathstrut_{\rho\sig}
\eeq
and the corresponding Ricci identity is
\beq
\label{Ricci_id}
\nabla_\al \nabla_\be u_\ga - \nabla_\be \nabla_\al u_\ga =
R_{\al\be,\ga}\mathstrut^{\de} u_{\de}.
\eeq
(The comma in $R_{\al\be,\ga\de}$ is used to distinguish the two
antisymmetric index pairs $[\al\be]$ and $[\ga\de]$.)

The connection coefficients are written in 3+1 form as
\beqa
\label{connection}
\om^{i}\mathstrut_{jk} &=& \Gam^{i}\mathstrut_{jk} = 
\bar \Gam^{i}\mathstrut_{jk}, \\
\om^{i}\mathstrut_{j0} &=& - N K^{i}\mathstrut_j + \d_j \be^i, \nonumber \\
\om^{i}\mathstrut_{0j} &=& - N K^{i}\mathstrut_j, \nonumber \\
\om^{i}\mathstrut_{00} &=& N g^{ij} \d_j N, \nonumber \\
\om^{0}\mathstrut_{0i} &=& \om^{0}\mathstrut_{i0} = N^{-1}  \d_i N, \nonumber \\
\om^{0}\mathstrut_{00} &=& N^{-1} \d_0 N, \nonumber 
\eeqa
and
\beq
\label{Kdef}
\om^{0}\mathstrut_{ij} = {1\over 2} N^{-2} \dzeroh g_{ij} = - N^{-1} K_{ij},
\eeq
where $K_{ij}$ is the extrinsic curvature (second fundamental tensor) of
$M_t$ and for any $t$-dependent space tensor ${\bf T}$, we define
another such tensor $\dzeroh {\bf T}$ of the same type by setting, as
in (\ref{Kdef}),
\beq
\dzeroh = {\d\over \d t} - {\cal L}_\be,
\eeq
where ${\cal L}_\be$ is the Lie derivative on $M_t$ with respect to
the spatial vector $\mathbf{\be}$.  Note that $\dzeroh$ and $\d_i=\d/\d x^i$
commute.

\section{Bianchi Equations}

\indent We recall that the Riemann tensor of a pseudo-Riemannian
metric satisfies the Bianchi identities
\beq
\label{Bianchi}
\nabla_\al R_{\be\ga,\la\mu} + \nabla_\ga R_{\al\be,\la\mu} +
\nabla_\be R_{\ga\al,\la\mu} \equiv 0.
\eeq
These identities imply by contraction and use of the symmetries
of the Riemann tensor
\beq
\label{Bianchi1c}
\nabla_\al R^\al\mathstrut_{\be,\la\mu} + \nabla_\mu R_{\la\be} 
-\nabla_\la R_{\mu\be} \equiv 0,
\eeq
where the Ricci tensor is defined by 
$$R^{\al}\mathstrut_{\be,\al\mu}=R_{\be\mu}.$$
If the Ricci tensor satisfies the Einstein equations
\beq
\label{Einstein}
R_{\al\be} = \rho_{\al\be},
\eeq
then the previous identities imply the equations
\beq
\label{Bianchi1c_mat}
\nabla_\al R^\al\mathstrut_{\be,\la\mu} = 
\nabla_\la \rho_{\mu\be} - \nabla_\mu \rho_{\la\be} .
\eeq
Equation (\ref{Bianchi}) with $\{\al\be\ga\} = \{ijk\}$ and
(\ref{Bianchi1c_mat}) with $\be=0$ do not contain derivatives
of the Riemann tensor transversal to $M_t$; hence, we consider
these equations as constraints (``Bianchi constraints'').  We
shall consider the remaining equations in (\ref{Bianchi}) and 
(\ref{Bianchi1c_mat}) as applying to a double two-form 
$A_{\al\be,\la\mu}$, which is simply a spacetime tensor
antisymmetric in its first and last pairs of indices.  We do
{\it not} suppose {\it a priori} a symmetry between the two
pairs of antisymmetric indices.  These equations, called from here
on ``Bianchi equations,'' read as follows
\beq
\label{Bianchi_eq1}
\nabla_0 A_{hk,\la\mu} + \nabla_k A_{0h,\la\mu} + 
\nabla_h A_{k0,\la\mu} = 0, 
\eeq
\beq
\label{Bianchi_eq2}
\nabla_0 A^0\mathstrut_{i,\la\mu} + \nabla_h A^h\mathstrut_{i,\la\mu}
= \nabla_\la \rho_{\mu i} - \nabla_\mu \rho_{\la i} \equiv
J_{\la\mu i}, 
\eeq
where the pair $[\la\mu]$ is either $[0j]$ or $[jl]$.  We next
introduce\cite{cite7} two ``electric'' and two ``magnetic'' space
tensors associated with the double two-form $\bA$, in analogy
to the electric and magnetic vectors associated with the electromagnetic
two-form $\bF$.  That is, we define the ``electric'' tensors by
\beq
\label{electric}
E_{ij} \equiv A^0\mathstrut_{i,0j} = -N^{-2} A_{0i,0j}, 
\eeq
$$
D_{ij} \equiv {1\over 4} \eta_{ihk} \eta_{jlm} A^{hk,lm}, 
$$
while the ``magnetic'' tensors are given by
\beqa
\label{magnetic}
H_{ij} \equiv {1\over 2} N^{-1} \eta_{ihk} A^{hk}\mathstrut_{,0j}, \\
B_{ji} \equiv {1\over 2} N^{-1} \eta_{ihk} A_{0j,}\mathstrut^{hk}.
\nonumber
\eeqa
In these formulae, $\eta_{ijk}$ is the volume form of the space metric
$\barbg$.
\vspace{0.2cm}

\noindent {\bf Lemma.} 
{\it (1) If the double two-form $\bA$ is symmetric with respect to its 
two pairs of
antisymmetric indices, then $E_{ij}=E_{ji}$, $D_{ij}=D_{ji}$, and
$H_{ij}=B_{ji}$. (2) If $\bA$ is a symmetric double two-form such that
\beq
A_{\al\be} \equiv A^{\la}\mathstrut_{\al,\la\be} = c g_{\al\be},
\eeq
then $H_{ij}=H_{ji}=B_{ji}=B_{ij}$ and $E_{ij}=D_{ij}$. }
\vspace{0.2cm}

{\it Proof.} (1) If $\bA$ is a {\it symmetric} double two-form, the proof is
immediate.  (2) The Lanczos identity \cite{cite8} for a symmetric double 
two-form, with a tilde representing the spacetime double dual, is given by
\beq
\tilde A_{\al\be,\la\mu} + A_{\al\be,\la\mu} \equiv C_{\al\la} g_{\be\mu}
-C_{\al\mu} g_{\be\la} + C_{\be\mu} g_{\al\la} - C_{\be\la} g_{\al \mu},
\eeq
with
\beqa
C_{\al\be} &\equiv& A_{\al\be} - {1\over 4} A g_{\al\be}, \\
A &\equiv& A^{\la\al}\mathstrut_{,\la\al}. \nonumber
\eeqa
The equalities $\bE=\bD$ and $\bB=\bH$ follow by a simple calculation that
employs the relation $\eta_{0ijk}=N\eta_{ijk}$ between the spacetime and
space volume forms.
\vspace{0.2cm}

In order to extend the treatment to the non-vacuum case and to avoid 
introducing
unphysical characteristics in the solution of the Bianchi equations, we will
keep as independent unknowns the four tensors $\bE$, $\bD$, $\bB$, and $\bH$,
which will not be regarded necessarily as symmetric.  The symmetries will be
imposed eventually on the initial data and shown to be conserved by
evolution.

We now express the Bianchi equations in terms of the time dependent space
tensors $\bE$, $\bH$, $\bD$, and $\bB$.  We use the following relations
found by inverting the equations (\ref{electric}) and (\ref{magnetic})
\beqa
\label{Eij}
A_{0i,0j} &=& -N^2 E_{ij}, \\
\label{Hij}
A_{hk,0j} &=& N \eta^i\mathstrut_{hk} H_{ij}, \\
\label{Dij}
A_{hk,lm} &=& \eta^i\mathstrut_{hk} \eta^j\mathstrut_{lm} D_{ij}, \\
\label{Bij}
A_{0j,hk} &=& N \eta^i\mathstrut_{hk} B_{ji}.
\eeqa
We will express spacetime covariant derivatives $\nabla$ of the spacetime
tensor $\bA$ in terms of space covariant derivatives $\bgrad$ and
time derivatives $\dzeroh$ of $\bE$, $\bH$, $\bD$, and $\bB$ by using
the connection coefficients in 3+1 form as given in (\ref{connection})
and (\ref{Kdef}).

The first Bianchi equation (\ref{Bianchi_eq1}) with $[\la\mu]=[0j]$
has the form
\beq
\nabla_0 A_{hk,0j} + \nabla_k A_{0h,0j} -\nabla_h A_{0k,0j} = 0.
\eeq
A calculation incorporating (\ref{connection}), (\ref{Kdef}) and
(\ref{Eij})-(\ref{Bij}), then grouping derivatives using 
$\dzeroh$ and $\bgrad_i$, yields the first Bianchi equation in the
form
\beq
\label{dzeroH}
\dzeroh (\eta^i\mathstrut_{hk} H_{ij}) + 2 N \bgrad_{[h} E_{k]j} +
(L_1)_{hk,j} = 0, 
\eeq
\beqa
\label{L1}
(L_1)_{hk,j} &\equiv&
 N K^l\mathstrut_j \eta^i\mathstrut_{hk} H_{il}
+ 2 (\bgrad_{[h} N) E_{k]j}
+ 2 N \eta^i\mathstrut_{lj} K^l\mathstrut_{[k}B_{h]i} \\
&&\hspace{1cm}
-(\bgrad^l N) \eta^i\mathstrut_{hk} \eta^m\mathstrut_{lj} D_{im}.
\nonumber
\eeqa
The second Bianchi equation (\ref{Bianchi_eq2}), with $[\la\mu]=[0j]$,
has the form
\beq
\nabla_0 A^0\mathstrut_{i,0j} + \nabla_h A^h\mathstrut_{i,0j} = J_{0ji},
\eeq
where $\bJ$ is zero in vacuum.  A calculation similar to the one above
yields for the second Bianchi equation
\beq
\label{dzeroE}
\dzeroh E_{ij} - N \eta^{hl}\mathstrut_i \bgrad_h H_{lj} + (L_2)_{ij} =
J_{0ji},
\eeq
\beqa
\label{L2}
(L_2)_{ij} &\equiv& -N(\tr \bK) E_{ij} + N K^k\mathstrut_j E_{ik}
+2 N K_i\mathstrut^k E_{kj} \\
&&\hspace{1cm}  
- (\bgrad_h N) \eta^{hl}\mathstrut_i H_{lj}
+ N K^k\mathstrut_h \eta^{lh}\mathstrut_i \eta^{m}\mathstrut_{kj} D_{lm}
+ (\bgrad^k N) \eta^l\mathstrut_{kj} B_{il}. \nonumber
\eeqa

The non-principal terms in the first two Bianchi equations  (\ref{dzeroH})
and (\ref{dzeroE}) are linear in $\bE$, $\bD$, $\bB$, and $\bH$, as well
as in the other geometrical elements $N\bK$ and $\bbgrad N$.  The
characteristic matrix of the principal terms is symmetrizable.  The
unknowns $E_{i(j)}$ and $H_{i(j)}$, with fixed $j$ and $i=1,2,3$ appear
only in the equations with given $j$.  The other unknowns appear in 
non-principal terms.  The characteristic matrix is composed of three
blocks around the diagonal, each corresponding to one given $j$.

The $j^{\rm th}$ block of the characteristic matrix in an orthonormal
frame for the space metric $\barbg$, with unknowns listed horizontally
and equations listed vertically, ($j$ is suppressed) is given by
\beq
\label{matrix}
\begin{array}{cc}
\begin{array}{c}

\end{array}
&
\begin{array}{cccccc}
\ \ E_1\ \  & \ \ E_2\ \  & \ \ E_3\ \  & \ \ H_1\ \  & 
\ \ H_2\ \  & \ \ H_3\ \ 
\end{array}
\\
\begin{array}{c}
(\ref{dzeroE})_1 \\
(\ref{dzeroE})_2 \\
(\ref{dzeroE})_3 \\
(\ref{dzeroH})_{23} \\
(\ref{dzeroH})_{31} \\
(\ref{dzeroH})_{12}
\end{array}
&
\left(
\begin{array}{cccccc}
\xi_0 & 0 & 0 & 0 & N \xi_3 & - N \xi_2 \\
0 & \xi_0 & 0 & - N \xi_3 & 0 & N \xi_1 \\
0 & 0 & \xi_0 & N \xi_2 & - N \xi_1 & 0 \\
0 & - N \xi_3 & N \xi_2 & \xi_0 & 0 & 0 \\
N \xi_3 & 0 & - N \xi_1 & 0 & \xi_0 & 0 \\
- N \xi_2 & N \xi_1 & 0 & 0 & 0 & \xi_0 
\end{array}
\right) .
\end{array}
\eeq
This matrix is symmetric and its determinant is the characteristic polynomial 
of the $\bE$, $\bH$ system.  It is given by 
\beq
-N^6 (\xi_0 \xi^0) (\xi_\al \xi^\al)^2.
\eeq
The characteristic matrix is symmetric in an orthonormal space frame 
and the timelike direction defined by $\dzeroh$ has a coefficient
matrix $T_0$ that is positive definite (here $T_0$ is the unit matrix).
Therefore, the first order system is symmetrizable hyperbolic.  We do 
not have to compute the symmetrized form explicitly because we will obtain the
energy estimate directly by using the contravariant associates
$E^{ij}$, $H^{ij}, \ldots$ of the unknowns.

The second pair of Bianchi equations is obtained from (\ref{Bianchi_eq1})
and (\ref{Bianchi_eq2}) with $[\la\mu]=[lm]$.  We obtain from 
(\ref{Bianchi_eq1})
\beq
\label{dzeroD}
\dzeroh (\eta^i\mathstrut_{hk} \eta^j\mathstrut_{lm} D_{ij} )
+ 2 N \eta^j\mathstrut_{lm} \bgrad_{[k} B_{h]j} +
(L_3)_{hk,lm} = 0,
\eeq
\beqa
\label{L3}
(L_3)_{hk,lm} &\equiv& 
2 N \eta^n\mathstrut_{j[m} K^j\mathstrut_{l]} \eta^i\mathstrut_{hk} D_{in} 
+ 2 \eta^j\mathstrut_{lm} (\bgrad_{[k} N) B_{h]j} \\
&&\hspace{1cm} 
+ 2 N K_{l[h} E_{k]m} + 2 N K_{m[k} E_{h]l}
+ 2 H_{i[l} ( \bgrad_{m]} N ) \eta^i\mathstrut_{hk}. \nonumber
\eeqa
Analogously, from (\ref{Bianchi_eq2}) we obtain
\beq
\label{dzeroB}
\dzeroh (\eta^j\mathstrut_{lm} B_{ij}) 
- N \eta^{kh}\mathstrut_i  \eta^j\mathstrut_{lm} \bgrad_h D_{kj}
+ (L_4)_{i,lm} = -N J_{lmi},
\eeq
\beqa
\label{L4}
(L_4)_{i,lm} &\equiv& 
-N (\tr \bK) \eta^j\mathstrut_{lm} B_{ij}
+ 2 N \eta^h\mathstrut_{j[m} K^j\mathstrut_{l]} B_{ih}
+ 2 N K^h\mathstrut_i \eta^j\mathstrut_{lm} B_{hj} \\
&&\hspace{1cm} 
- (\bgrad_j N) \eta^{hj}\mathstrut_i  \eta^n\mathstrut_{lm} D_{hn}
- 2 N \eta^j\mathstrut_{hi} H_{j[m} K^h\mathstrut_{l]} 
+ 2 E_{i[m} \bgrad_{l]} N
.\nonumber 
\eeqa

Consider the system (\ref{dzeroD}) and (\ref{dzeroB}) with $[lm]$
fixed.  Then $j$ in $\eta_{jlm}$ is also fixed.  The characteristic
matrix for the $[lm]$ equations, with unknowns $D_{ij}$ and $B_{ij}$,
$j$ fixed, with an orthonormal space frame, is the same as the matrix
(\ref{matrix}).

If the spacetime metric $\bg$ is considered as given, as well as the
sources, the Bianchi equations (\ref{dzeroH}), (\ref{dzeroE}), (\ref{dzeroD}),
and (\ref{dzeroB}) form a linear symmetric hyperbolic system with domain of
dependence determined by the light cone of $\bg$.  The coefficients of the
terms of order zero are $\bbgrad N$ or $N\bK$.  The system is homogeneous
in vacuum (zero sources).

\section{Bel Energy in a Strip}

Multiply (\ref{dzeroH}) by ${1\over 2}\eta_l\mathstrut^{hk} H^{lj}$ and
recall that 
$\eta_l\mathstrut^{hk}\eta^i\mathstrut_{hk} = 2 \delta^i\mathstrut_l$, 
$\eta_{lrk}\eta^{ihk} = \delta^i\mathstrut_l \delta^h\mathstrut_r-
\delta^i\mathstrut_r \delta^h\mathstrut_l$, and
$\dzeroh g^{ij} = 2 N K^{ij}$.  Then we find
that
\beq
\label{dzeroHsq}
{1\over 2}\eta_l\mathstrut^{hk} H^{lj} \dzeroh(\eta^i\mathstrut_{hk} H_{ij})
= {1\over 2} \dzeroh(H_{ij} H^{ij}) - M_1,
\eeq
\beqa
\label{M1}
M_1 &\equiv& {1\over 4} \eta^l\mathstrut_{rs}H_{lm} \eta^i\mathstrut_{hk} 
H_{ij} \dzeroh ( g^{hr} g^{ks} g^{jm} ) \\
&=& N ((\tr \bK) H^{ij} -K^i\mathstrut_l H^{lj} 
+ K_l\mathstrut^j H^{il}) H_{ij}. 
\nonumber
\eeqa

Likewise, multiply (\ref{dzeroE}) by $E^{ij}$ to obtain
\beq
\label{dzeroEsq}
E^{ij} \dzeroh E_{ij} = {1\over 2} \dzeroh ( E_{ij} E^{ij}) - M_2,
\eeq
\beq
\label{M2}
M_2 \equiv N (K^i\mathstrut_l E^{lj}  +
 K_l\mathstrut^{j} E^{il}) E_{ij}.
\eeq
Multiply (\ref{dzeroD}) by $(1/4)\eta_r\mathstrut^{hk} \eta_s\mathstrut^{lm}
D^{rs}$ and (\ref{dzeroB}) by $(1/2) \eta_h\mathstrut^{lm} B^{ih}$ to
obtain analogous results for the second pair of Bianchi equations.  Sum the
expressions so obtained from the four Bianchi equations (\ref{dzeroH}),
(\ref{dzeroE}), (\ref{dzeroD}), and (\ref{dzeroB}).  The spatial derivatives
add to form an exact spatial divergence, just as for all symmetric systems.
Indeed, we obtain
\beqa
\label{dzeroSum}
{1\over 2} \dzeroh ( |\bE|^2 + |\bH|^2 + |\bD|^2 + |\bB|^2) +
\bgrad_h(N E^{ij} \eta^{lh}\mathstrut_i H_{lj}) && \\
&&\hspace{-5.5cm}- \bgrad_h (N B^{ij} \eta^{lh}\mathstrut_i D_{lj}) =
Q(\bE,\bH,\bD,\bB) + {\cal S}, \nonumber 
\eeqa
where we have denoted by $|\cdot|$ the pointwise $\barbg$ norm of a space
tensor, and where $Q$ is a quadratic form with coefficients
$\bbgrad N$ and $N\bK$ given by
\beqa
\label{Qdef}
Q &\equiv& -{1\over 2} \eta_l\mathstrut^{hk} H^{lj} (L_1)_{hk,j} 
-E^{ij} (L_2)_{ij} 
-{1\over 4} \eta_r\mathstrut^{hk} \eta_s\mathstrut^{lm} D^{rs} (L_3)_{hk,lm} \\
&& -{1\over 2} \eta_h\mathstrut^{lm} B^{ih} (L_4)_{i,lm}
+(\bgrad_h N) E^{ij} \eta^{lh}\mathstrut_i H_{lj}
-(\bgrad_h N) B^{ij} \eta^{lh}\mathstrut_i D_{lj} \nonumber \\
&& +M_1 + M_2 + M_3 + M_4. \nonumber
\eeqa
The source term ${\cal S}$, zero in vacuum, is
\beq
\label{source}
{\cal S} \equiv J_{0ij} E^{ij} 
- {1\over 2} N J_{lmi} \eta_h\mathstrut^{lm} B^{ih}.
\eeq

We define the {\it Bel energy} at time $t$ of the field $(\bE,\bH,\bD,\bB)$,
called a ``Bianchi field'' when it satisfies the Bianchi equations, to be
the integral
\beq
\label{BelEnergy}
{\cal B}(t) \equiv {1\over 2} \int_{M_t} ( |\bE|^2 + |\bH|^2 + |\bD|^2 + 
|\bB|^2) \mu_{\barbg_t}.
\eeq

We prove the following
\vspace{.2cm}

\noindent {\bf Theorem 1.} {\it Suppose that $\bg$ is ${\cal C}^1$ on 
$M\times[0,T]$
and that the $\barbg_t$ norms of $\bbgrad N$ and $N \bK$ are uniformly
bounded on $M_t$, $t\in [0,T]$.  Denote by $\pi(t)$ the supremum
\beq
\label{pi}
\pi(t) = \Sup_{M_t} (|\bbgrad N| + |N\bK|).
\eeq
Suppose the matter source $\bJ\in L^1([0,T],L^2(M_t))$, then the Bel
energy of a ${\cal C}^1$ Bianchi field with compact support in space
satisfies for $0\le t\le T$ the following inequality
\beq
\label{BelIneq1}
{\cal B}(t) \le \biggl\{ {\cal B}(0) +
 \int_0^t \| {\cal S} \|_{L^2(M_\tau)} d\tau \biggr\} 
\exp( C \int_0^t \pi(\tau)d\tau),
\eeq
where $C$ is a given positive number. }
\vspace{.2cm}

\noindent {\it Proof.}  We integrate the identity (\ref{dzeroSum}) on the strip
$M\times [0,t]$ with respect to the volume element $\mu_{\barbg_\tau}d\tau$.
If the Bianchi field has support compact in space, the integral of the
space divergence term vanishes.  The integration of a function $\dzeroh F$
on a strip of spacetime with respect to the volume form 
$d\tau\,\mu_{\barbg_\tau}$ goes as follows
\beqa
\label{integ}
\int_0^t \int_{M_\tau} \dzeroh F \, d\tau\,\mu_{\barbg_{\tau}} &\equiv&
\int_0^t \int_{M_\tau} (\d_t - \beta^i \d_i) F \,d\tau\,\mu_{\barbg_{\tau}} 
\nonumber\\
&=& \int_0^t \int_{M_\tau} \d_\tau ( F (\det \barbg_\tau)^{1/2} ) \,d\tau 
\,d^3 x \\
&&- \int_0^t \int_{M_\tau} \biggl( F  (\det \barbg_\tau)^{-1/2} \d_\tau 
(\det \barbg_\tau)^{1/2} \biggr) \,\mu_{\barbg_\tau}\, d\tau \nonumber \\ 
&& - \int_0^t \int_{M_\tau} \bgrad_i(\beta^i F) 
\,\mu_{\barbg_\tau}\, d\tau + 
\int_0^t \int_{M_\tau} F \bgrad_i \beta^i \, \mu_{\barbg_\tau}\, d\tau.
\nonumber
\eeqa
Therefore, if $F$ has compact support in space, we can
express the right hand side of (\ref{integ}) as
\beq
\label{integ2}
\int_{M_\tau} F \, \mu_{\barbg_\tau} -
\int_{M_0} F \, \mu_{\barbg_0} +
\int_0^t \int_{M_\tau} N (\tr \bK) F \, \mu_{\barbg_\tau}\, d\tau,
\eeq
where we have used
\beq
\d_\tau  (\det \barbg_\tau)^{1/2} = -  (\det \barbg_\tau)^{1/2}
N (\tr \bK) +  (\det \barbg_\tau)^{1/2} \bgrad_i \beta^i
\eeq
with $\tr \bK= K^j\mathstrut_j$.

The integration of (\ref{dzeroSum}) on a strip leads therefore to the
equality
\beq
{\cal B}(t) = {\cal B}(0) + \int_0^t \int_{M_\tau} (\tilde Q +
{\cal S})  \, \mu_{\barbg_\tau}\, d\tau,
\eeq
with 
$$\tilde Q = Q + 
{1\over 2} N(\tr \bK) ( |\bE|^2 + |\bH|^2 + |\bD|^2 + |\bB|^2).$$
We deduce from this equality and the expression (\ref{Qdef})
for $Q$ the following inequality, with $C$ some number
\beq
\label{BelIneq2}
{\cal B}(t) \le {\cal B}(0) + 
C\int_0^t \Sup_{M_\tau} ( | \bbgrad N | + | N \bK |) {\cal B}(\tau)
\, d\tau + 
\int_0^t \| {\cal S} \|_{L^2(M_\tau)} \,d\tau.
\eeq
This inequality and the resolution of the corresponding equality
imply the result.
\vspace{.2cm}

\noindent {\bf Corollary.} {\it If the metrics $\barbg_t$ are complete, 
the inequality
given in the theorem extends to ${\cal C}^1$ Bianchi fields that
are square integrable in $\mu_{\barbg_t}$.}
\vspace{.2cm}

\noindent {\it Proof.}  One uses a truncating sequence.
\vspace{.2cm}

\noindent {\bf Remark 1.}  The quantities $-\bgrad_k N$ and $-2 K_{ij}$
are, respectively, the $(0k)$ and $(ij)$ components of the Lie
derivative ${\cal L}_{\bf n} {\bf g}$ of the spacetime metric ${\bf g}$
with respect to the unit normal ${\bf n}$ of $M_t$.   [Its
$(00)$ component is identically zero.]  The Bel energy, 
therefore, is conserved if this Lie derivative is zero.
\vspace{.2cm}

\noindent {\bf Remark 2.} Energy estimates for the double two-form $\bA$ can
be deduced from the Bel-Robinson tensor of $\bA$ when $\bA$ is assumed
to be symmetric in its pairs of indices and when a timelike vector
field has been chosen\cite{cite8b}.  However, we do not make such a hypothesis
about the symmetry of $\bA$ here.

\section{Local Energy Estimate}

We take as a domain $\Omega$ of spacetime the closure of a connected
open set whose boundary $\d\Omega$ consists of three parts: a domain
$\om_t$ of $M_t$, a domain $\om_0$ of $M_0$, and a lateral boundary
$L$.  We assume $L$ is spacelike or null and ``ingoing,''  that is,
timelike lines entering $\Omega$ at a point of $L$ are past directed.
We also assume that the boundary $\d\Omega$ is regular in the
sense of Stoke's formula.  We use the identity (\ref{dzeroSum})
previously found and integrate this identity on $\Omega$ with
respect to the volume form $ \mu_{\barbg_\tau}\, d\tau$.  Let
$f(t,x) = t + \phi(x)$ be the local equation of $L$.  Set $\nu_i=
\d_i f = \d_i \phi$ and $\nu_0 = 1 - \be^i \nu_i$; these are the
components in the coframe $\th^\al$ of the spacetime gradient
$\d_\al f= \nu_\al$.  Then we have from (\ref{dzeroSum})
\beqa
\label{integL}
{1\over 2} \int_{\om_t} ( |\bE|^2 + |\bH|^2 + |\bD|^2 + |\bB|^2)
\, \mu_{\barbg_t} 
-{1\over 2} \int_{\om_0} ( |\bE|^2 + |\bH|^2 + |\bD|^2 + |\bB|^2)
\, \mu_{\barbg_0} && \nonumber \\
+ \int_L \biggl[ {1\over 2}\nu_0 ( |\bE|^2 + |\bH|^2 + |\bD|^2 + |\bB|^2)
+ N \nu_h \bigl\{ (\bE\wedge \bH)^h - (\bB \wedge \bD)^h \bigr\} \biggr]
\mu_{\barbg} &=& \nonumber \\
&&\hspace{-6.5cm}= 
\int_0^t \int_{\om_\tau} (\tilde Q + {\cal S}) \,\mu_{\barbg_\tau}\, d\tau.
\eeqa
We have set
$$  (\bE\wedge \bH)^h = \eta^{lh}\mathstrut_i E^{ij} H_{lj} \equiv
\sum_j (\bE\wedge \bH)^h_{(j)},$$
that is, $(\bE\wedge \bH)^h_{(j)}$, for each fixed $j$, is the
vector product in three dimensions of the vectors $\vec E^{(j)}$
and $\vec H_{(j)}$.   Therefore, the $\barbg$ norm of the
vector $(\bE\wedge \bH)^h$ satisfies
\beq
\label{EwHnorm}
| \bE\wedge \bH | \le \sum_j | \vec E^{(j)} \wedge \vec H_{(j)} |
\le \sum_j |\vec E^{(j)} | |\vec H_{(j)} | \le 
{1\over 2} ( | \bE |^2 + | \bH |^2).
\eeq

Let $\bar\nu= (\nu_i)$, $\nu_0 >0$, and 
$|\bar\nu|_{\barbg} \le N^{-1} \nu_0$, that is, let $\underline\nu=
(\nu_0,\nu_i)$ be timelike or null, as it must be for $L$.  [Note
that our sign conventions for $\underline\nu$ are suitable for
applications of Stokes's theorem in a spacetime with signature
$(-+++)$.] We can now deduce from (\ref{EwHnorm}) and an analogous
inequality for $(\bD,\bB)$ that the integral over $L$ in
(\ref{integL}) is non-negative.  It is strictly positive if
$\underline\nu$ is timelike.
\vspace{.2cm}

\noindent {\bf Remark.}  The integral over a null boundary $L$ is also strictly
positive if (\ref{EwHnorm}) is a strict inequality.  This will be
the case if any of the norms of the vector products of $\vec E^{(j)}$
by $\vec H_{(j)}$ is less than the product of the norms of
these vectors, that is, if any of these pairs of vectors are
non-orthogonal:  $E^{i(j)} H_{i(j)} \ne 0$ or $D^{i(j)} B_{i(j)} \ne 0$
for some $j$.  An analogous property has been obtained through the
use of the Bel-Robinson tensor of the Weyl curvature in \cite{cite9}.
\vspace{.2cm}

We define ${\cal B}(\om_t)$ by replacing in the definition of ${\cal B}(t)$
the integral over $M_t$ by an integral over $\om_t$
\beq
{\cal B}(\om_t) = {1\over 2} \int_{\om_t} 
( |\bE|^2 + |\bH|^2 + |\bD|^2 + |\bB|^2)\, \mu_{\barbg_t} .
\eeq
A result of the non-negativity of the integral on $L$ is that the
inequality (\ref{BelIneq2}) holds on $\Omega$ with ${\cal B}(\tau)$
replaced by ${\cal B}(\om_\tau)$, $\om_\tau \equiv M_\tau \cap \Omega$,
and $M_\tau$ replaced by $\om_\tau$.  Therefore the inequality
(\ref{BelIneq1}) also holds with the same replacements.  In particular,
we have ${\cal B}(\om_\tau)=0$ if ${\cal B}(\om_0)=0$ and $\bJ=0$
(vacuum case).  Then $\bE=\bH=\bD=\bB=0$ in $\Omega$ if they vanish
on the intersection of $M_0$ with the past of $\Omega$.  (Cf.
for related results \cite{cite9} and \cite{cite10}.)  Note that
such a result is not sufficient to prove the propagation of
gravitation with the speed of light because it treats only 
curvature tensors that are zero in some domain, not the difference of
non-zero curvature tensors.  The Bianchi equations are not by 
themselves sufficient to estimate such differences because their
coefficients depend on the metric, which itself depends on
the curvature.

In the next section, we will give a further first order symmetric hyperbolic
system linking the metric and the connection to our Bianchi
field.  This further system is inspired by an analogous one constructed
in conjunction with the Weyl tensor by Friedrich \cite{cite2}.

\section{Determination of ($\mathbf{\bar\Gamma,K}$) from Knowledge
of the Bianchi Fields}

We will need the $3+1$ decomposition of the Riemann tensor, which is
found by combining (\ref{Riemann}), (\ref{connection}), and
(\ref{Kdef}); namely,
\beqa
\label{Rijkl}
R_{ij,kl} &=& \bar R_{ij,kl} + 2 K_{i[k} K_{l]j}, \\
\label{R0ijk}
R_{0i,jk} &=& 2 N \bgrad_{[j} K_{k]i}, \\
\label{R0i0j}
R_{0i,0j} &=& N (\dzeroh K_{ij} + N K_{ik} K^{k}\mathstrut_j +
\bgrad_i \d_j N).
\eeqa
From these formulae one obtains the following ones for the Ricci
curvature $R_{\al\be} = R^{\ga}\mathstrut_{\al,\ga\be}$
\beqa
\label{Rij}
R_{ij} &=& \bar R_{ij} - N^{-1} \dzeroh K_{ij} + K_{ij} \tr \bK
-2 K_{ik} K^k\mathstrut_j - N^{-1} \bgrad_i \d_j N, \\
\label{R0j}
R_{0j} &=& N (\d_j \tr\bK - \bgrad_h K^h\mathstrut_j), \\
\label{R00}
R_{00} &=& N (\dzeroh \tr\bK - N K_{ij} K^{ij} + \bgrad^i \d_i N).
\eeqa
The identity
\beq
\label{dg}
\dzeroh g_{ij} \equiv -2 N K_{ij}
\eeq
and the expression for the spatial Christoffel symbols give
\beq
\label{dGam}
\dzeroh \bar\Gamma^h\mathstrut_{ij} \equiv
\bgrad^h(N K_{ij}) - 2 \bgrad_{(i}( N K_{j)}\mathstrut^h).
\eeq
Therefore, from the identity (\ref{R0ijk}), we obtain the
identity
\beq
\label{GamEq}
\dzeroh \bar\Gamma^h\mathstrut_{ij} + N \bgrad^h K_{ij} =
K_{ij} \d^h N - 2 K^h\mathstrut_{(i} \d_{j)} N -
2 R_{0(i,j)}\mathstrut^h.
\eeq
On the other hand, the identities (\ref{R0i0j}) and (\ref{Rij})
imply the identity
\beq
\label{KEq}
\dzeroh K_{ij} + N \bar R_{ij} + \bgrad_j \d_i N \equiv
-2 N R^{0}\mathstrut_{i,0j} - N (\trK ) K_{ij} + N R_{ij}.
\eeq
We obtain equations relating $\bbGam$ and $\bK$ to a
double two-form $\bA$ and matter sources by replacing, in the identities
(\ref{GamEq}) and (\ref{KEq}), $R_{0(i,j)}\mathstrut^h$ by
$(A_{0(i,j)}\mathstrut^h + A^h\mathstrut_{(j,i)0})/2$, 
$R^0\mathstrut_{i,0j}$ by
$(A^0\mathstrut_{i,0j} + A^0\mathstrut_{j,0i})/2$, and the
Ricci tensor of spacetime by a given tensor $\brho$, zero in
vacuum.  The terms involving $\bA$ are then replaced by Bianchi fields
as in (\ref{Eij})-(\ref{Bij}).

The first set of identities (\ref{GamEq}) leads to equations with
principal terms
\beq
\dzeroh \bGam^h\mathstrut_{ij} + N g^{hk} \d_k K_{ij}.
\eeq
To deduce from the second identity (\ref{KEq}) equations which will
form together with the previous ones a symmetric hyperbolic system, we
set
\beq
\label{Ndef}
N= \al^{-1} (\det \barbg)^{1/2},
\eeq
where $\al$ is a given scalar density of weight one.  This is the
``algebraic gauge'' \cite{cite11}.  (It can also be considered
as a change in the name of the unknown from $N$ to $\al$ \cite{cite12}.)
The condition (\ref{Ndef}), if $\bbGam$ denotes the
Christoffel symbols of $\barbg$, implies that
\beq
\bGam^h\mathstrut_{ih} = \d_i \log N + \d_i \log \al.
\eeq

The second set of identities (\ref{KEq}) now yields the following
equations, where $N$ denotes $\al^{-1} (\det \barbg)^{1/2}$,
\beqa
\label{KEq2}
\dzeroh K_{ij} + N \d_h \bGam^{h}\mathstrut_{ij} &=& N \bigl[
\bGam^m\mathstrut_{ih} \bGam^{h}\mathstrut_{jm} -
(\bGam^h\mathstrut_{ih} - \d_i \log \al) (\bGam^k\mathstrut_{jk}
-\d_j \log \al) \bigr] \\
&&\hspace{-2.5cm} + N (\d_i\d_j \log \al - \bGam^k\mathstrut_{ij}
\d_k \log \al) - N (E_{ij} + E_{ji})- N (\trK ) K_{ij} + N \rho_{ij}. \nonumber
\eeqa
The first set (\ref{GamEq}) yields
\beqa
\label{GamEq2}
\dzeroh \bGam^h\mathstrut_{ij} + N \bgrad^h K_{ij} &=&
N K_{ij} g^{hk} ( \bGam^m\mathstrut_{mk} - \d_k \log \al) \\
&&\hspace{-1cm}
-2 N K^h\mathstrut_{(i}( \bGam^m\mathstrut_{j)m} - \d_{j)} \log \al)
-N \eta^k\mathstrut_{(i}\mathstrut^h B_{j) k} 
- N H_{k(j} \eta^k\mathstrut_{i)}\mathstrut^h. \nonumber
\eeqa

We see from the principal parts of (\ref{KEq2}) and (\ref{GamEq2})
that the system obtained for $\bK$ and $\bbGam$ has
a characteristic matrix composed of six blocks around the diagonal,
each block a four-by-four matrix that is symmetrizable hyperbolic
with characteristic polynomial $N^4(\xi_0\xi^0)(\xi_\al \xi^\al)$.
The characteristic matrix in a spatial orthonormal frame has blocks
of the form
\beq
\left(
\begin{array}{cccc}
\xi_0 & N \xi_1 & N \xi_2 & N \xi_3 \\
N \xi_1 & \xi_0 & 0 & 0 \\
N \xi_2 & 0 & \xi_0 & 0 \\
N \xi_3 & 0 & 0 & \xi_0 
\end{array}
\right).
\eeq
\vspace{0.2cm}

{\noindent}{\bf Remark.}  In the equations considered above, $N$ is to be
replaced by $\al^{-1} (\det \barbg)^{1/2}$.  Instead of this
algebraic replacement, we can consider an evolution equation for $N$
of the type
\beq
\label{harmonic}
\dzeroh N + N^2 \trK = N f,
\eeq
where $f$ is an arbitrary function on spacetime.  For $f=0$, 
(\ref{harmonic}) is the harmonic slicing condition.  For arbitrary
$f$, the general solution of (\ref{harmonic}), if $\barbg$ and
$\bK$ are linked as in (\ref{dg}), is
\beq
\label{Ngam}
N = \gam^{-1} (\det\barbg)^{1/2},
\eeq
with $\gam$ such that
\beq
\label{harmonic2}
\dzeroh \gam + f\gam =0,
\eeq
that is, $\gam$ satisfies a linear first order differential equation
depending only on the known quantities ${\mathbf{\beta}}$ and $f$.
It does not depend on any of the previously defined unknowns
$(\bE,\bH,\bD,\bB,\barbg,\bK,\bbGam)$ of our system in algebraic
gauge.  The modified system consists of the equations in algebraic
gauge [(\ref{dzeroH}), (\ref{dzeroE}), (\ref{dzeroD}), (\ref{dzeroB}),
(\ref{dg}), (\ref{KEq2}), (\ref{GamEq2})] but with $N$ now replaced
by (\ref{Ngam}).  Therefore, $\al$ is replaced by $\gam$.  The
additional equation is (\ref{harmonic}), or, on account of (\ref{Ngam}),
this role can be regarded as played by (\ref{harmonic2}).

\section{Symmetric Hyperbolic System for ${\bf u}\equiv
(\bE,\bH,\\
\bD,\bB,\barbg,\bK,\bbGam)$}

We denote by $\bS$ the system composed of the equations 
(\ref{dzeroH}), (\ref{dzeroE}), (\ref{dzeroD}), (\ref{dzeroB}),
(\ref{dg}), (\ref{KEq2}), and (\ref{GamEq2}), where the lapse
function $N$ is replaced by $\al^{-1}(\det\barbg)^{1/2}$.  This
system is satisfied by solutions of the Einstein equations whose
shift $\bbeta$, hidden in the operator $\dzeroh$, has
the given arbitrary values and whose lapse has the form
$N=\al^{-1} (\det\barbg)^{1/2}$.  (Clearly, any $N>0$ can be
written in this form.)  The following lemma results from the
previous paragraphs.
\vspace{0.2cm}

\noindent {\bf Lemma.} {\it For arbitrary $\al$ and $\bbeta$, and
given matter sources $\brho$, the system $\bS$ is a first order
symmetric hyperbolic system for the unknowns $\bf u$. }
\vspace{0.2cm}

Note that in this Lemma, the various elements $\bE$, $\bH$, $\bD$,
$\bB$, $\barbg$, $\bK$, and $\bbGam$ are considered as independent.
For example, {\it a priori}, we neither know that $\bbGam$ denotes
the Christoffel symbols of $\barbg$ nor that $\bE,\ \bH,\ \bD$,
and $\bB$ are identified with components of the Riemann tensor
of spacetime.

An analogous Lemma holds for the system $\bS'$ that has an
additional unknown $N$ (equivalently, $\gam$) and an additional
equation (\ref{harmonic}) [equivalently, (\ref{harmonic2})]
with $f$ replacing $\al$ as an arbitrary function.

We now consider the {\it vacuum} case.
\vspace{0.2cm}

\noindent{\bf Initial Values.}  The original Cauchy data for
the Einstein equations are, with $\bphi$ a properly
Riemannian metric and $\bpsi$ a second rank tensor on
$M_0$,
\beq
\barbg|_0 = \bphi,\quad \bK|_0 = \bpsi.
\eeq
The tensors $\bphi$ and $\bpsi$ must satisfy the constraints, which
read in vacuum,
\beq
R_{0i} = 0,  
\eeq
\beqa
0 = G_{00} &\equiv& R_{00} - {1\over 2} g_{00} R \\
&=& R_{00} + {1\over 2} N^2 g^{\al\be} R_{\al\be} \nonumber \\
&=&{1\over 2} N^2 ( \bar R - K_{ij} K^{ij} + (\trK)^2), \nonumber
\eeqa
with $\bar R= g^{ij} \bar R_{ij}$.  The initial data given by
$\bphi$ determine the Cauchy data $\bGam^{h}\mathstrut_{ij}|_0$ and
thus $\bar R_{ij,kl}|_0$.  Then, $R_{ij,kl}|_0$
is determined by using also $\bpsi$.  To determine the initial
values of the other components of the unknown $\bf u$ of the
system $\bS$, we use the arbitrarily given data $\al$ and
$\bbeta$.  In particular, we use $N= \al^{-1} (\det\barbg)^{1/2}$
to find $R_{0i,jk}|_0$ and to compute $\bgrad_j\d_i N|_0$
appearing in the identity (\ref{Rij}).  We deduce from
(\ref{Rij}) $\dzeroh K_{ij}|_0$ when $R_{ij}=0$, which enables
$R_{0i,0j}$ to be found from (\ref{R0i0j}).  All of the
components of the Riemann tensor of spacetime are then known
on $M_0$.  We identify them with the corresponding components of
the double two-form $\bA$ on $M_0$:  the latter have thus
{\it initially} the same symmetries as the Riemann tensor.  We
find the initial values of $(\bE,\bH,\bD,\bB)$ according to their
definitions in terms of $\bA$.

\section{Existence Theorems}

In order to define Sobolev spaces on a manifold $M$,
it is convenient to endow $M$ with a ${\cal C}^\infty$
Riemannian metric $\bfe$.  We will suppose that $\bfe$ has
a non-zero injectivity radius, $\delta_0>0$, hence is complete,
and that it has Riemannian curvature uniformly bounded on $M$,
as well as its derivatives of all orders relevant for the
required Sobolev properties to hold.

With $I$ an interval of $R$, and under the hypotheses made on
$\bfe$, each manifold $M\times I$ endowed with the metric
$\bfe - dt^2$ is globally hyperbolic \cite{cite13,cite14}.
We denote by $D$ the covariant derivative associated with $\bfe$
(the ``$\bfe$-covariant derivative'').
\vspace{0.2cm}

\noindent {\bf Theorem 2.} 
{\it Hypotheses. a.  The arbitrary quantities
are such that, with $I$ an interval containing $0$, $\al$ is
continuous on $M\times I$, and there exist numbers $A_1,\ A_2>0$
such that on $M\times I$
$$ 0< A_1 \le \al \le A_2, $$
$$ {\mathbf{D}}\al,\bbeta \in {\cal C}^0(I,H_{s+1}).$$
b. The initial data are such that
$$ {\mathbf{D}}\bphi,\bpsi \in H_{s+1},$$
and $\bphi$ is a continuous properly Riemannian metric on $M$ 
uniformly equivalent to $\bfe$.  That is, there exist strictly
positive numbers $b_1$, $b_2$ such that at each point of $M$
and for any tangent vector $\xi$ to $M$ at this point,
$$ b_1 \bfe(\bxi,\bxi) \le \bphi(\bxi,\bxi) \le b_2 \bfe(\bxi,\bxi).$$}

\noindent {\it Conclusion. If $s\ge 3$, there exists a number $T>0$,
$[0,T]\in I$, such that the system $\bS$ has one and only one
${\cal C}^1$ solution $\bf u$ on $[0,T]\times M$ taking the
Cauchy data ${\bf u}_0$ deduced from $\bphi,\ \bpsi, \al, \bbeta$.
The solution $\bf u$ is such that the components of $\bf u$
different from $\barbg$, as well as $\barbg-\bphi$, are in
${\cal C}^0([0,T],H_s)\cap {\cal C}^1([0,T],H_{s-1}).$
There exist $B_1$ and $B_2$ such that
$$ B_1 \bfe(\bxi,\bxi) \le \barbg (\bxi,\bxi) \le B_2 \bfe(\bxi,\bxi).$$}
\vspace{0.2cm}

\noindent {\it Proof.}  1. We replace the equation (\ref{dg}) in
the system $\bS$ by the equation
\beq
\label{dgphi}
\dzeroh (g_{ij} - \phi_{ij}) = -2 N K_{ij} + {\cal L}_\beta \phi_{ij}.
\eeq
We still have a symmetric first order system $\tilde\bS$ for the
unknown $\bf\tilde u$, deduced from $\bf u$ by replacing $\barbg$
by $\barbg-\bphi$.  This quasilinear system is hyperbolic as
long as $\barbg$ is properly Riemannian.

The hypotheses made on the arbitrary data $\al,\ \bbeta$ and on the
initial data $\bphi,\ \bpsi$ imply, through the Sobolev
multiplication properties, that the initial data ${\bf \tilde u}|_0$
belong to $H_s$ whenever $s\ge 1$.
\vspace{0.2cm}

\noindent 2.  If the Riemannian manifold $(M,\bfe)$ is the
Euclidean space $R^3$, the existence theorem is known,
with $s> 3/2 +1$. [Linear case: Friedrichs \cite{cite15}.  Quasilinear
case: Fischer-Marsden \cite{cite16}, using semigroup methods, and
Majda \cite{cite17}, using energy estimates.]

A proof under the hypotheses we have made on $(M,\bfe)$ and on the
data can be obtained along the same lines, using energy estimates,
because the spaces $H_s$ on $(M,\bfe)$ have the same functional
properties that they have on $R^3$.  The energy inequalities for 
$\bf \tilde u$ and its spatially covariant $\bfe$-derivatives of
order $\le s$ can be obtained directly on $[0,T]\times M$,
as we have indicated in the case of the Bianchi equations.  One uses
$s$-energy estimates for linear symmetric first order hyperbolic
systems and the G\aa rding duality method \cite{cite18} to
prove existence and uniqueness for the linearized system.  One
uses the contraction mapping principle in the norm
${\cal C}^0([0,T],H_{s-1})$ on the one hand, and boundedness
in the norm ${\cal C}^0([0,T],H_s) \cap {\cal C}^1([0,T],H_{s-1})$
on the other hand, as in Majda \cite{cite17}, to prove the theorem
for our quasilinear system.
\vspace{0.2cm}

\noindent {\bf Remark.}  The spacetime defined by the $\barbg$ component
of $\bf u$ together with $\al$ and $\bbeta$ on $M\times [0,T]$
is globally hyperbolic, with a ${\cal C}^1$ metric if $\al$ and
$\bbeta$ are ${\cal C}^1$.
\vspace{0.2cm}

\noindent {\bf Domain of dependence.}  The exterior sheet of the 
characteristic cone of the system $\bS$ at a point of spacetime
$V$ is the light cone of the spacetime metric $\bg$.  It
has been proven by Leray \cite{cite13}, for globally hyperbolic
Leray systems on a manifold, that the values of a solution of the
Cauchy problem in a compact set $\bar\om_t \subset M_t$ depend
only on the values of the Cauchy data in the compact set $\bar\om_0$
that is the intersection of $M_0$ with the past of $\bar\om_t$.
We will prove this result for our symmetric hyperbolic system.

We recall that on a spacetime $(V,\bg)$, $V=M\times R$, temporally
oriented by the orientation of $R$, the past $\cP(x)$ of a
point $x$ is the union of the paths timelike with respect to
$\bg$ and ending at $x$.  The past $\cP(\om)$ of a
subset $\om\subset V$ is the union of the past of its points.
\vspace{0.2cm}

\noindent {\bf Lemma.} {\it Let $(V,\bg)$ be a globally hyperbolic
manifold with $V= M\times R$ and $\bg$ a ${\cal C}^1$ metric.
Let $\bar\om_T$ be a compact subset of $M_T$ with non-empty
interior $\om_T$ and boundary $\d \om_T$. Then,
$$ \bar\Omega = \cP(\bar\om_T) \cap \{ t\ge0\} $$
is a compact subset of $V$.  Its boundary is
$$ \d \Omega \equiv \bar\om_T \cup \bar\om_0 \cup L. $$
The lateral boundary $L$ is generated by null geodesic arcs
issuing from points of $\d \om_T$. At an interior point of such
an arc, $L$ has a null tangent hyperplane and is ingoing.  }
\vspace{0.2cm}

\noindent {\it Proof.}  On a globally hyperbolic manifold, the
past of a closed set is closed;  and the intersection of the past
of a compact set with a ``past compact set'' (here, $t\ge 0$)
is compact.  Therefore, $\bar\Omega$ is compact.

If $y\in \bar\Omega\cap \{0<t<T\}$, there exists $x\in\bar\om_T$
such that $y\in \cP(x)$.  Suppose that the path $xy$ is 
strictly timelike.  Then there is an open neighborhood $V(y)$
of $y$ in $\cP(x)$ such that $z\in V(y)$ implies that $z\in \cP(x)$.
Consequently, $y$ cannot be on the boundary $L$, which must
therefore be generated by null paths.  Analogous reasoning
shows that $x$ cannot be an interior point of $\om_T$.  For
the proof of the rest of this Lemma, see Leray \cite{cite13},
Lemma 97.
\vspace{0.2cm}

Recognizing that in a $\cC^1$ globally hyperbolic manifold the
lateral boundary $L$ admits a null tangent
hyperplane almost everywhere, we prove the following theorem.
\vspace{0.2cm}

\noindent {\bf Theorem 3.} {\it (Domain of Dependence)  Two $\cC^1$
solutions $\bu_1$ and $\bu_2$ of $\bS$ coincide on $\om_T$ if
the arbitrary data $\al,\ \bbeta$ coincide in the past of $\om_T$
defined by $\bg_1$, while the Cauchy data $(\bphi_1,\bpsi_1)$
and $(\bphi_2,\bpsi_2)$ coincide on $\om_0$, which is the
intersection of the $\bg_1$-past of $\om_T$ with $M_0$.}
\vspace{0.2cm}

\noindent {\it Proof.}  The difference of two solutions satisfies
a linear homogeneous first order symmetric hyperbolic system
with $\cC^1$ principal coefficients and $\cC^0$ non-principal
ones.  We set $\bgam= \bbGam_1 - \bbGam_2$, $\bk = \bK_1 - \bK_2$,
$\bG = \barbg_1 - \barbg_2$, $\ba = \bA_1 -\bA_2 $, and likewise for
the Bianchi fields obtained from $\bA$. The 
linear system is composed of three sets of equations in which we
denote by $\bell$ various linear forms in $\bgam,\ \bk,\ldots$ with
coefficients analytic in $\bu_1$ and $\bu_2$.

The first set is
\beqa
\label{dgam1}
\dzeroh \gamma^h\mathstrut_{ij} + N_1 g^{hk}_1 \d_k k_{ij} 
&=& \ell^{(\gam)}_{ij}, \\
\label{dk1}
\dzeroh k_{ij} + N_1 \d_h \gam^h\mathstrut_{ij} &=&
\ell^{(k)}_{ij}. 
\eeqa
The second set consists of an equation for $\bbe = \bE_1 - \bE_2$
\beq
\label{de}
\dzeroh e_{ij} - N_1 g_1^{hm} g_1^{ln} \eta_{1mni} h_{lj} = 
\ell^{(e)}_{ij}
\eeq
and similar equations for $\bh= \bH_1 - \bH_2$, $\bd = \bD_1 - \bD_2$,
and $\bb = \bB_1 - \bB_2$ deduced from the Bianchi equations.
[Note that $\bbe$ in (\ref{de}) is not to be confused with the
metric $\bfe$ introduced in Sec. 8 and below in the discussion of
$H^{{\mathrm{u.l.}}}_s$.]  The third set of equations is
\beq
\label{dG}
\dzeroh G_{ij} = \ell^{(G)}_{ij}.
\eeq

Raising and lowering indices with the metric $\barbg_1$, we
multiply (\ref{dgam1}) by $\gam_h\mathstrut^{ij}$, (\ref{dk1})
by $k^{ij}$, (\ref{de}) by $e^{ij}$ and so on.  We add the
results and integrate over a domain $\Omega$ of the type
indicated in the preceding Lemma, constructed with $\bg_1$.
We denote by $\cE(\om_t)$ the integral
\beq
\label{EEnergy}
\cE(\om_t) = {1\over 2} \int_{\om_t} ( | \bgam |^2 + | \bk |^2
+ | \bbe |^2 + | \bh |^2 + | \bd |^2 + | \bb | ^2 + | \bG |^2)
\mu_{\barbg_1}.
\eeq
Integration of (\ref{EEnergy}) over $\Omega$ leads to the
identity
\beq
\cE(\om_T) = \cE(\om_0) + \int_L \bsig + \int_0^T \int_{\om_t}
q \mu_{\barbg_1(t)} dt,
\eeq
where $\bsig$ is a three-form and $q$ is a quadratic function of
$\bgam,\ \bk,\ldots, \bG$ with coefficients analytic in
$\bu_1$ and $\bu_2$.  The three-form $\bsig$ arises from integration
of a $\dzeroh$ derivative together with a spatial divergence from
the principal terms of our equations.  We made an analogous
computation for the Bianchi equations and showed that the integral
over $L$ in that case is positive.  The same argument holds here for
the terms coming from the second set of equations.  It is
easy to prove that the integral over $L$ is also positive for
the first and third sets of equations.  On the other hand,
$q$ is bounded on $\Omega$, up to a multiplicative constant $C$,
by the sum of the squares of the $\barbg_1$ norms of $\bgam,\ 
\bk,\ \bbe,\ldots,\ \bG$.  Hence, we have the inequality
\beq
\cE(\om_T) \le \cE(\om_0) + C \int_0^T \cE (\om_t) dt.
\eeq
It follows that $\cE(\om_T) = 0 $ if $\cE(\om_0)=0$, which
is the stated result.
\vspace{0.2cm}

By using Theorem 3, we will show that the local existence result
holds for a non-compact manifold $M$ without any hypotheses on
the fall-off at infinity of the initial data.
\vspace{0.2cm}

\noindent {\bf Definition.}  {\it A tensor $\bu$ on $M$ belongs to
a space $H^{{\mathrm{u.l.}}}_s$ (``u.l.'' denotes ``uniformly local'')
if there exists a covering $\bC$ of $M$ by open sets $\om_{(i)}$
with compact closures such that
\begin{enumerate}
\item The restriction of $\bu$ to each of these open sets belongs to
$H_s(\om_{(i)})$, that is, $\bu$ is square-integrable on $\om_{(i)}$
in the volume element defined  by the metric $\bfe$ introduced in 
Sec. 8, and so are all its $\bfe$-covariant derivatives of order $\le s$.

\item $$ \Sup_i \| \bu \| _{H_s(\om_{(i)} )} \equiv
\| \bu \|_{H^{{\mathrm{u.l.}}}_s} < \infty.$$ 
\end{enumerate}
}
\vspace{0.2cm}

Note that different choices of uniformly locally finite coverings, as well 
as different uniformly equivalent metrics $\bfe$, define the 
same $H^{{\mathrm{u.l.}}}_s$ space because they define equivalent 
norms.
\vspace{0.2cm}

\noindent {\bf Lemma.} {\it If, for the covering $\bC$, the arbitrary
quantities and the original initial data are such that
$$ \al \in \cC^0(I,H^{{\mathrm{u.l.}}}_{s+2}), \quad \al>0, $$
$$ \bbeta\in \cC^0(I,H^{{\mathrm{u.l.}}}_{s+1}),$$
$$ \bphi \in H^{{\mathrm{u.l.}}}_{s+2},\quad 
\bpsi \in H^{{\mathrm{u.l.}}}_{s+1},$$
and $\bphi$ defines a properly Riemannian metric on $M$, uniformly
equivalent to $\bfe$, then the initial data for $\bu$ belong
to $H^{{\mathrm{u.l.}}}_s$ if $s\ge 1$. }
\vspace{0.2cm}

\noindent {\it Proof.} The proof follows from the definition and
the Sobolev multiplication properties.
\vspace{0.2cm}

\noindent {\bf Theorem 4.}  
{\it Hypotheses. a. The arbitrary quantities are such that,
with $I$ an interval containing $0$,
$$ \al \in \cC^0(I,H^{{\mathrm{u.l.}}}_{s+2}), \quad \al>A >0, $$
$$ \bbeta\in \cC^0(I,H^{{\mathrm{u.l.}}}_{s+1}).$$

\noindent b. The initial data are such that
$$ \bphi \in H^{{\mathrm{u.l.}}}_{s+2},\quad 
\bpsi \in H^{{\mathrm{u.l.}}}_{s+1},$$
and $\bphi$ defines a properly Riemannian metric on $M$, uniformly
equivalent to $\bfe$.

\noindent Conclusion.  If $s\ge 3$, there exists a number $T>0$,
$[0,T]\subset I$, such that the system $\bS$ has one and only one
solution $\bu\in \cC^0([0,T],H^{{\mathrm{u.l.}}}_{s}) \cap
\cC^1([0,T],H^{{\mathrm{u.l.}}}_{s-1})$ taking the Cauchy data
$\bu_0 \in H^{{\mathrm{u.l.}}}_{s}$ deduced from $\bphi,\ \bpsi$. }
\vspace{0.2cm}

\noindent {\it Proof.} Suppose $M$ is not compact.  Consider a covering
of $M_T$ by compact sets $K_T^{(i)}$.  The pasts of these sets in
the globally hyperbolic metric $ -A^2 dt^2 + B_2 \bfe$ 
cover $V_T \equiv M\times [0,T]$.  (The numbers $A$ and $B_2$
here refer to Theorem 2.)  The intersections of these pasts with
$M_0$ define a covering of $M_0$ by  compact sets $K^{(i)}_0$.
There exist truncating functions $\theta^{(i)}$ that are smooth,
bounded uniformly with respect to $i$ (as are their $\bfe$-covariant
derivatives of order $\le s$), equal to one on $K^{(i)}_0$ and
equal to zero outside another compact set $\tilde K^{(i)}$
strictly containing $K^{(i)}_0$ \cite{cite19}.  Use the
sets $\tilde K^{(i)}$ to define the $H^{{\mathrm{u.l.}}}_{s}$
norms of tensors on $M$.

Consider the Cauchy problem for the system $\tilde\bS$ with Cauchy data
$\tilde \bu_0^{(i)} = \theta^{(i)} \tilde\bu_0$.  There exists
a constant $C$ independent of $i$ such that
$$ \| \tilde \bu_0^{(i)} \|_{H_s} \le 
C \| \tilde \bu_0\|_{H^{{\mathrm{u.l.}}}_{s}}. $$
Corresponding inequalities hold for the given quantities $\al,\ \bbeta$.
Theorem 2 shows that there exists a number $T'<T$, independent of $i$,
such that $\bS$ has one and only one solution $\bu^{(i)}$ on
$M\times [0,T]$ that takes the initial value $\tilde \bu_0^{(i)}$.
\vspace{0.2cm}

\noindent {\bf Remark.}  The associated metric $\barbg^{(i)}$ is
uniformly equivalent to $\bfe$ and the spacetime metric $\bg^{(i)}$
satisfies the inequalities posited in Theorem 2.  Therefore, the past
of a subset in this metric is contained in the past defined with
$\bfe$.
\vspace{0.2cm}

We obtain, from the set of $u^{(i)}$'s defined on $V_T$,
a solution $\bu$ for the Cauchy problem for $\bS$ on $V_t$ by
setting
$$ \bu(x,t) = u^{(i)}(x,t)\ {\mathrm{on}}\ \cP(K^{(i)}_T)\cap
\{0\le t\le T'\} $$
because, by Theorem 3 and the previous Remark, we have 
$$ u^{(j)}(x,t) = u^{(i)}(x,t)\ {\mathrm{on}}\ \cP(K^{(j)}_T)\cap
\cP(K^{(i)}_T)\cap \{0\le t\le T'\}.$$

Finally, we prove
\vspace{0.2cm}

\noindent {\bf Theorem 5.} {\it The solution of $\bS$ with the data
deduced from the arbitrary quantities $\al,\ \bbeta$ and the
initial data $\bphi,\ \bpsi$ is a solution of the vacuum Einstein
equations with lapse $N= \al^{-1} (\det \barbg)^{1/2}$ and
shift $\bbeta$ if the initial data satisfy the constraints.}
\vspace{0.2cm}

\noindent {\it Proof.} The result holds for $H_{s+1}$ as well
as for $H^{{\mathrm{u.l.}}}_{s+1}$ data.  We spell it out for the
first case.  It has been proven by Choquet-Bruhat and York\cite{cite4},
for initial data in the indicated function spaces and satisfying
the constraints, and for a given arbitrary shift $\bbeta$, that the
vacuum Einstein equations have one and only one solution in the
``algebraic gauge,'' that is, with a lapse function in the given
form, such that
$\barbg \in \cC^0(M\times [0,T]),\ (\bD \barbg,\bK) \in
\cC^0([0,T],H_{s+1})$.  The connection of $\barbg$, the second
fundamental form $\bK$ and the Riemann tensor of $\barbg$ satisfy
the system $\bS$ and take the Cauchy data $\bu_0$.  They coincide,
therefore, with the solution $\bu$ found previously.  In particular,
the component $\barbg$ of $\bu$ together with 
$N= \al^{-1} (\det\barbg)^{1/2}$ and $\bbeta$ are a metric that satisfies
the vacuum Einstein equations.
\vspace{0.2cm}

\noindent {\bf Remark.} It is a result of the coincidence of the
components $\barbg$ and $\bK$ of $\bu$ with the solution of the
Einstein equations with initial data $\bphi$ and $\bpsi$, given
shift $\bbeta$, and lapse $N= \al^{-1} (\det\barbg)^{1/2}$ that
these components are in fact such that $(\bD \barbg,\bK) \in
\cC^0([0,T],H_{s+1})$.
\vspace{1cm}

Acknowledgment.  AA and JWY thank the National Science Foundation
of the USA, grants PHY 94-13207 and PHY 93-18152/ASC 93-18152 (ARPA 
supplemented) for support.  JWY thanks Michael
E. Taylor for a useful discussion.

\end{document}